\documentclass[aps,prd,11pt,superscriptaddress,amssymb,amsmath,showpacs,nofootinbib]{revtex4-1}
\pdfoutput=1
\usepackage{hhline}
\usepackage{color}
\usepackage{epsfig}
\usepackage{graphicx}
\usepackage{float}
\usepackage{hyperref}
\usepackage{epstopdf}
\usepackage[usenames,dvipsnames]{xcolor}
\usepackage{multirow,booktabs}
\usepackage[normalem]{ulem}
\usepackage[d]{esvect}
\usepackage[misc,geometry]{ifsym}
\usepackage{harpoon}

\begin{document}

\title{ Connect discreteness to continuousness in leptonic flavor symmetries }

\author{Ding-Hui Xu}\email{3036602895@qq.com}
\author{Shu-Jun Rong\Letter}\email{rongshj@glut.edu.cn}
\affiliation{College of Physics and Electronic Information Engineering, Guilin University of Technology, Guilin, Guangxi 541004, China}

\begin{abstract}
Discrete groups are widely used in the expression of flavor symmetries of leptons. In this paper, we employ
a novel mathematical object called group-algebra (GA) to describe symmetries of the leptonic mass matrices.
A GA element is constructed by a discrete group with continuous parameters.  For a GA element, there is an equivalent symmetry which can be continuous, discrete, and hybrid. According to the equivalence between a GA element and other symmetries, we perform a classification of 198 nontrivial elements of the GA  generated
by the group $S_{4}$. Based on the results of the classification, the phenomenological consequences of the $S_{4}$ GA are illustrated.

\end{abstract}
	

\maketitle
	
\section{Introduction}
As is known, the success of the standard model (SM) of particle physics is based on the Lorentz symmetry and the gauge symmetries.
Nature chooses the symmetries as the architecture principle of the physical theory. Following the belief of the principle, some complicated and novel
symmetries are proposed to postulate new physics beyond the SM. For example, in the filed of neutrino physics, flavor symmetries (FSs)\cite{Yanagida:1980xy,Ohlsson:2002rb,Babu:2002dz,Ma:2004zv,Grimus:2005rf,Lam:2006wm,Xing:2006xa,He:2006dk,Altarelli:2009gn,Chen:2009um,Antusch:2011qg,Hernandez:2012ra,Feruglio:2017spp}, also called horizontal symmetries, are
introduced. In contrast to the gauge symmetry, FSs are used to explore possible patterns hidden in the leptonic mixing parameters. In general, these symmetries are expressed by
finite discrete groups. For reviews, see refs\cite{Altarelli:2010gt,King:2013eh,Xing:2015fdg,Feruglio:2019ybq,Xing:2020ijf,Ding:2023htn,Ding:2024ozt}. The multiplication rules and the representation of a flavor group can determine the texture of the leptonic mass matrix and the mixing pattern.  However, because the correspondence between the symmetry and the values of the mixing parameters is not unique, one may wander in the endless list of the candidates
of symmetries. Maybe, a critical question should  be taken into account, namely, whether there is a
continuous symmetry leading to the same mass matrix as what predicted by a discrete flavor group? In other words, the question whether there is a more general structure beyond finite groups to describe flavor symmetries deserves a consideration.
	
One of the authors of this paper introduced a novel mathematical structure called group-algebra (GA) to express the symmetry of a mass matrix\cite{Rong:2017rel,Rong:2019qjh}.
The element of a GA is a linear superposition of elements of a group. Apparently, a GA shows symmetries beyond the extent of the group as the generator. However, there are interesting relationships between a discrete group and a continuous GA. The previous work\cite{Rong:2019qjh} shows that the elements of the $S_{3}$ GA are equivalent to the  symmetry  $Z_{2}\times CP$, where $CP$ denotes a generalised CP transformation\cite{Grimus:2003yn,Feruglio:2012cw,Feruglio:2013hia,Holthausen:2012dk,Ding:2013bpa,Rong:2016cpk,Chen:2018lsv,Novichkov:2019sqv,Nilles:2020nnc}. A discrete symmetry is connected to a continuous one in the case of GA. Furthermore, both the discrete and continuous symmetry can lead to the same mass matrix. Thus, we may consider that Nature has more choices in the expression of symmetries besides groups.
To check the observation in more general cases, we perform a complete classification of elements of the 2-dimensional $S_{4}$ GA and study the relationships between the elements and $Z_{2}\times CP$ (or other symmetries). On the basis of the work, we may propose more general symmetry realizations which can fit the reported leptonic mixing parameters.

The outline of the paper is as follows. In Sec. \uppercase\expandafter{\romannumeral2}, the definition and the representation of 2-dimensional GA is recapitulated.
 In Sec. \uppercase\expandafter{\romannumeral3}, the classification of the elements of $S_{4}$ GA are conducted. The relationships between the the GA elements and $Z_{2}\times CP$
 are studied. The phenomenological consequences of the $S_{4}$ GA are also discussed in this section. Finally, a conclusion is given.

\section{Definition and representation of a 2-dimensional GA}
A GA K[G] is constructed by the linear superposition of elements of the group G with coefficients in the filed K, i.e.\cite{Rong:2019qjh},
\begin{equation}
\label{eq:1}
 \sum_{g\in G} a_{g}g .
\end{equation}
K[G] is an algebra on the bases of the addition and the multiplication rules as follows,
\begin{equation}\label{eq:2}
\begin{array}{c}
 \sum_{g\in G} a_{g}g+\sum_{g\in G} b_{g}g=\sum_{g\in G}( a_{g}+b_{g})g, \\
 (\sum_{g\in G} a_{g}g)(\sum_{h\in G} b_{h}h)=\sum_{g\in G, h\in G}( a_{g}b_{h})g h.
  \end{array}
 \end{equation}

For simplicity, we consider a 2-dimensional GA expressed as follow
\begin{equation}
\label{eq:3}
X=x_{1}A_{1}+x_{2}A_{2},
\end{equation}
with $A_{1}$, $A_{2}$, being elements of a finite discrete group.
If $X$ is used to express a symmetry in the vector space of a physical theory, its representation should be unitary, namely
\begin{equation}
\label{eq:4}
X^{+}X=XX^{+}=I,
\end{equation}
where $I$ is the unit matrix.
According to the unitary condition, a 2-dimensional GA can be realized in the following manner\cite{Rong:2019qjh}
\begin{equation}\label{eq:5}
X(\theta)=\cos\theta A_{1}+i\sin\theta A_{2}.
\end{equation}
Here i is the imaginary factor, $A_{1}$ and $A_{2}$ satisfy the constraints
\begin{equation}
\label{eq:6}
A_{1}A_{2}^{+}=A_{2}A_{1}^{+},~~ A_{1}^{+}A_{2}=A_{2}^{+}A_{1}.
\end{equation}
Thus, they have the property  $(A_{1}A_{2}^{+})^{2}=(A_{1}^{+}A_{2})^{2}=I$.
Furthermore, if we define $B\equiv A^{+}_{1}A_{2}$, $X$ can be rewritten as\cite{Rong:2019qjh}
\begin{equation}\label{eq:7}
X(\theta)=A_{1}e^{i\theta B}.
\end{equation}
Then we can view a 2-dimensional GA as a complex-number-like generalization in the field of matrix, namely $i^{2}B^{2}=-I$.

Let us give some comments on the properties of $X({\theta})$. First, although $A_{i}$ is a discrete group-element, $X(\theta)$ itself is a structure describing a continuous
symmetry with the parameter $\theta$. Second, since in general cases $X(\theta)^{2}\neq I$, it cannot be employed to express the residual symmetry of Majorna neutrinos.
In other words, we can use $X(\theta)$ in the expression of symmetries of charged leptons or Dirac neutrinos.

\section{2-dimensional GA C[$S_{4}$]}
\subsection{$S_{4}$ group}
We consider the 2-dimensional GA based on the group $S_{4}$, denoted as C[$S_{4}$].
Its representation is derived directly from  the  representation of $S_{4}$. For the phenomenological purpose,
we concern the 3-dimensional representation. $S_{4}$ has two nonequivalent 3-dimensional representations. In the representation $\mathbf{3}$, the group elements are expressed as follows\cite{Ishimori:2010au}:
\begin{equation}
\label{eq:8}
\begin{array}{c}
  a_{1}=\left(
                 \begin{array}{ccc}
                   1 & 0 & 0 \\
                   0 & 1 &0 \\
                   0 & 0 & 1 \\
                 \end{array}
               \right),a_{2}=\left(
                 \begin{array}{ccc}
                   1 & 0 & 0 \\
                   0 & -1 & 0 \\
                   0 &0 & -1\\
                 \end{array}
               \right),a_{3}=\left(
                 \begin{array}{ccc}
                   -1 & 0 & 0 \\
                   0 & 1 & 0 \\
                   0 &0 & -1\\
                 \end{array}
               \right),a_{4}=\left(
                 \begin{array}{ccc}
                   -1 & 0 & 0 \\
                   0 & -1 & 0 \\
                   0 &0 & 1\\
                 \end{array}
               \right),\\
   b_{1}=\left(
                 \begin{array}{ccc}
                   0 & 0 & 1 \\
                   1 & 0 &0 \\
                   0 & 1 & 0 \\
                 \end{array}
               \right), b_{2}=\left(
                 \begin{array}{ccc}
                   0 & 0 & 1 \\
                   -1 & 0 &0 \\
                   0 & -1 & 0 \\
                 \end{array}
               \right), b_{3}=\left(
                 \begin{array}{ccc}
                   0 & 0 & -1 \\
                   1 & 0 &0 \\
                   0 & -1 & 0 \\
                 \end{array}
               \right), b_{4}=\left(
                 \begin{array}{ccc}
                   0 & 0 & -1 \\
                   -1 & 0 &0 \\
                   0 & 1 & 0 \\
                 \end{array}
               \right),
               \\
   c_{1}=\left(
                 \begin{array}{ccc}
                   0 & 1 & 0 \\
                   0 & 0 &1 \\
                   1& 0 & 0 \\
                 \end{array}
               \right), c_{2}=\left(
                 \begin{array}{ccc}
                   0 & 1 & 0 \\
                   0 & 0 &-1 \\
                   -1& 0 & 0 \\
                 \end{array}
               \right), c_{3}=\left(
                 \begin{array}{ccc}
                   0 & -1 & 0 \\
                   0 & 0 &1 \\
                   -1& 0 & 0 \\
                 \end{array}
               \right),  c_{4}=\left(
                 \begin{array}{ccc}
                   0 & -1 & 0 \\
                   0 & 0 &-1 \\
                   1& 0 & 0 \\
                 \end{array}
               \right),
                \\
   d_{1}=\left(
                 \begin{array}{ccc}
                  1 & 0 & 0 \\
                   0 & 0 &1 \\
                   0& 1 & 0 \\
                 \end{array}
               \right),  d_{2}=\left(
                 \begin{array}{ccc}
                  1 & 0 & 0 \\
                   0 & 0 &-1 \\
                   0& -1 & 0 \\
                 \end{array}
               \right), d_{3}=\left(
                 \begin{array}{ccc}
                  -1 & 0 & 0 \\
                   0 & 0 &1 \\
                   0& -1 & 0 \\
                 \end{array}
               \right), d_{4}=\left(
                 \begin{array}{ccc}
                  -1 & 0 & 0 \\
                   0 & 0 &-1 \\
                   0& 1 & 0 \\
                 \end{array}
               \right),
               \\
   e_{1}=\left(
                 \begin{array}{ccc}
                  0 & 1 & 0 \\
                   1 & 0 &0 \\
                   0& 0 & 1 \\
                 \end{array}
               \right),  e_{2}=\left(
                 \begin{array}{ccc}
                  0 & 1 & 0 \\
                   -1 & 0 &0 \\
                   0& 0 & -1 \\
                 \end{array}
               \right), e_{3}=\left(
                 \begin{array}{ccc}
                  0 & -1 & 0 \\
                   1 & 0 &0 \\
                   0& 0 & -1 \\
                 \end{array}
               \right),  e_{4}=\left(
                 \begin{array}{ccc}
                  0 & -1 & 0 \\
                   -1 & 0 &0 \\
                   0& 0 & 1 \\
                 \end{array}
               \right),
               \\
   f_{1}=\left(
                 \begin{array}{ccc}
                  0 & 0 & 1 \\
                   0 & 1 &0 \\
                  1& 0 & 0 \\
                 \end{array}
               \right), f_{2}=\left(
                 \begin{array}{ccc}
                  0 & 0 & 1 \\
                   0 & -1 &0 \\
                  -1& 0 & 0 \\
                 \end{array}
               \right), f_{3}=\left(
                 \begin{array}{ccc}
                  0 & 0 & -1 \\
                   0 & 1 &0 \\
                  -1& 0 & 0 \\
                 \end{array}
               \right), f_{4}=\left(
                 \begin{array}{ccc}
                  0 & 0 & -1 \\
                   0 & -1 &0 \\
                  1& 0 & 0 \\
                 \end{array}
               \right).
\end{array}
\end{equation}
For the other representation $\mathbf{3'}$, there are following relations\cite{Ishimori:2010au}: $a_{2}(\mathbf{3'})=a_{2}(\mathbf{3})$, $b_{2}(\mathbf{3'})=b_{2}(\mathbf{3})$,
$d_{1}(\mathbf{3'})=-d_{1}(\mathbf{3})$, $d_{3}(\mathbf{3'})=-d_{3}(\mathbf{3})$, $d_{4}(\mathbf{3'})=-d_{4}(\mathbf{3})$. However, according to the form of $X(\theta)$(Eq.\ref{eq:5}), we can see that  $\mathbf{3}$ and $\mathbf{3'}$ lead to the same GA representation. Hence, we just consider the representation $\mathbf{3}$ in the following sections.

\subsection{Elements of 2-dimensional C[$S_{4}$] }
There are 198 elements in total in the 2-dimensional GA C[$S_{4}$], which are constructed by 2 different non-identity elements of the group $S_{4}$. In the form of Eq.\ref{eq:7}, they are listed as follows:
\begin{equation}\label{eq:9}
\begin{array}{cccccc}
  X_{1}=a_{2}e^{i\theta a_{3}}, &  X_{2}=a_{2}e^{i\theta a_{4}},  &  X_{3}=a_{2}e^{i\theta d_{1}},  &  X_{4}=a_{2}e^{i\theta d_{2}},  &  X_{5}=a_{2}e^{i\theta e_{1}},  &  X_{6}=a_{2}e^{i\theta e_{4}},  \\
   X_{7}=a_{2}e^{i\theta f_{1}}, &  X_{8}=a_{2}e^{i\theta f_{3}},  &  X_{9}=a_{3}e^{i\theta a_{2}},  & X_{10}=a_{3}e^{i\theta a_{4}},  & X_{11}=a_{3}e^{i\theta d_{1}},  &  X_{12}=a_{3}e^{i\theta d_{2}} ,\\
   X_{13}=a_{3}e^{i\theta e_{1}}, & X_{14}=a_{3}e^{i\theta e_{4}},  &  X_{15}=a_{3}e^{i\theta f_{1}},  & X_{16}=a_{3}e^{i\theta f_{3}},  & X_{17}=a_{4}e^{i\theta a_{2}},  &  X_{18}=a_{4}e^{i\theta a_{3}}, \\
   X_{19}=a_{4}e^{i\theta d_{1}},&X_{20}=a_{4}e^{i\theta d_{2}},  & X_{21}=a_{4}e^{i\theta e_{1}},  & X_{22}=a_{4}e^{i\theta e_{4}},  & X_{23}=a_{4}e^{i\theta f_{1}},  &  X_{24}=a_{4}e^{i\theta f_{3}},\\
    X_{25}=d_{1}e^{i\theta a_{2}},& X_{26}=d_{1}e^{i\theta a_{3}},  &  X_{27}=d_{1}e^{i\theta a_{4}},  & X_{28}=d_{1}e^{i\theta d_{2}},  & X_{29}=d_{1}e^{i\theta e_{1}},  &  X_{30}=d_{1}e^{i\theta e_{4}},  \\
   X_{31}=d_{1}e^{i\theta f_{1}}, &  X_{32}=d_{1}e^{i\theta f_{3}},  & X_{33}=d_{2}e^{i\theta a_{2}},  & X_{34}=d_{2}e^{i\theta a_{3}}, &X_{35}=d_{2}e^{i\theta a_{4}},  &  X_{36}=d_{2}e^{i\theta d_{1}},\\
  X_{37}=d_{2}e^{i\theta e_{1}}, & X_{38}=d_{2}e^{i\theta e_{4}},  &  X_{39}=d_{2}e^{i\theta f_{1}},  & X_{40}=d_{2}e^{i\theta f_{3}},  & X_{41}=e_{1}e^{i\theta a_{2}},  &  X_{42}=e_{1}e^{i\theta a_{3}},  \\
 X_{43}=e_{1}e^{i\theta a_{4}}, &X_{44}=e_{1}e^{i\theta d_{1}},   &X_{45}=e_{1}e^{i\theta d_{2}},   &X_{46}=e_{1}e^{i\theta e_{4}},   & X_{47}=e_{1}e^{i\theta f_{1}},  &  X_{48}=e_{1}e^{i\theta f_{3}},\\
X_{49}=e_{4}e^{i\theta a_{2}}, &X_{50}=e_{4}e^{i\theta a_{3}},    &X_{51}=e_{4}e^{i\theta a_{4}},   &X_{52}=e_{4}e^{i\theta d_{1}},    & X_{53}=e_{4}e^{i\theta d_{2}},   &  X_{54}=e_{4}e^{i\theta e_{1}},\\
X_{55}=e_{4}e^{i\theta f_{1}}, &X_{56}=e_{4}e^{i\theta f_{3}},   &X_{57}=f_{1}e^{i\theta a_{2}},  &X_{58}=f_{1}e^{i\theta a_{3}},   & X_{59}=f_{1}e^{i\theta a_{4}},   &  X_{60}=f_{1}e^{i\theta d_{1}},\\
X_{61}=f_{1}e^{i\theta d_{2}},&X_{62}=f_{1}e^{i\theta e_{1}}, &X_{63}=f_{1}e^{i\theta e_{4}}, &X_{64}=f_{1}e^{i\theta f_{3}},   &X_{65}=f_{3}e^{i\theta a_{2}},   &  X_{66}=f_{3}e^{i\theta a_{3}},\\
X_{67}=f_{3}e^{i\theta a_{4}},&X_{68}=f_{3}e^{i\theta d_{1}},&X_{69}=f_{3}e^{i\theta d_{2}}, &X_{70}=f_{3}e^{i\theta e_{1}},   &X_{71}=f_{3}e^{i\theta e_{4}},   &  X_{72}=f_{3}e^{i\theta f_{1}},\\
X_{73}=b_{1}e^{i\theta a_{2}},&X_{74}=b_{1}e^{i\theta a_{3}},&X_{75}=b_{1}e^{i\theta a_{4}}, &X_{76}=b_{1}e^{i\theta d_{1}},   &X_{77}=b_{1}e^{i\theta d_{2}},   &  X_{78}=b_{1}e^{i\theta e_{1}},\\
X_{79}=b_{1}e^{i\theta e_{4}},&X_{80}=b_{1}e^{i\theta f_{1}},&X_{81}=b_{1}e^{i\theta f_{3}}, &X_{82}=b_{2}e^{i\theta a_{2}},  &X_{83}=b_{2}e^{i\theta a_{3}},   &  X_{84}=b_{2}e^{i\theta a_{4}},\\
X_{85}=b_{2}e^{i\theta d_{1}},&X_{86}=b_{2}e^{i\theta d_{2}},&X_{87}=b_{2}e^{i\theta e_{1}},&X_{88}=b_{2}e^{i\theta e_{4}},  &X_{89}=b_{2}e^{i\theta f_{1}},   &  X_{90}=b_{2}e^{i\theta f_{3}},\\
X_{91}=b_{3}e^{i\theta a_{2}},&X_{92}=b_{3}e^{i\theta a_{3}},&X_{93}=b_{3}e^{i\theta a_{4}},&X_{94}=b_{3}e^{i\theta d_{1}},  &X_{95}=b_{3}e^{i\theta d_{2}},   &  X_{96}=b_{3}e^{i\theta e_{1}},\\
X_{97}=b_{3}e^{i\theta e_{4}},&X_{98}=b_{3}e^{i\theta f_{1}},&X_{99}=b_{3}e^{i\theta f_{3}},&X_{100}=b_{4}e^{i\theta a_{2}},  &X_{101}=b_{4}e^{i\theta a_{3}},   &  X_{102}=b_{4}e^{i\theta a_{4}}, \\
X_{103}=b_{4}e^{i\theta d_{1}},&X_{104}=b_{4}e^{i\theta d_{2}},&X_{105}=b_{4}e^{i\theta e_{1}},&X_{106}=b_{4}e^{i\theta e_{4}},  &X_{107}=b_{4}e^{i\theta f_{1}},   &  X_{108}=b_{4}e^{i\theta f_{3}},\\
X_{109}=c_{1}e^{i\theta a_{2}},&X_{110}=c_{1}e^{i\theta a_{3}},&X_{111}=c_{1}e^{i\theta a_{4}},&X_{112}=c_{1}e^{i\theta d_{1}},  &X_{113}=c_{1}e^{i\theta d_{2}},   &  X_{114}=c_{1}e^{i\theta e_{1}},\\
X_{115}=c_{1}e^{i\theta e_{4}},&X_{116}=c_{1}e^{i\theta f_{1}},&X_{117}=c_{1}e^{i\theta f_{3}},&X_{118}=c_{2}e^{i\theta a_{2}},  &X_{119}=c_{2}e^{i\theta a_{3}},   &  X_{120}=c_{2}e^{i\theta a_{4}},\\
X_{121}=c_{2}e^{i\theta d_{1}},&X_{122}=c_{2}e^{i\theta d_{2}},&X_{123}=c_{2}e^{i\theta e_{1}},&X_{124}=c_{2}e^{i\theta e_{4}}, &X_{125}=c_{2}e^{i\theta f_{1}},   &  X_{126}=c_{2}e^{i\theta f_{3}},\\
X_{127}=c_{3}e^{i\theta a_{2}},&X_{128}=c_{3}e^{i\theta a_{3}},&X_{129}=c_{3}e^{i\theta a_{4}},&X_{130}=c_{3}e^{i\theta d_{1}}, &X_{131}=c_{3}e^{i\theta d_{2}},   &  X_{132}=c_{3}e^{i\theta e_{1}},\\
X_{133}=c_{3}e^{i\theta e_{4}},&X_{134}=c_{3}e^{i\theta f_{1}},&X_{135}=c_{3}e^{i\theta f_{3}},&X_{136}=c_{4}e^{i\theta a_{2}}, &X_{137}=c_{4}e^{i\theta a_{3}},    &  X_{138}=c_{4}e^{i\theta a_{4}}, \\
X_{139}=c_{4}e^{i\theta d_{1}},&X_{140}=c_{4}e^{i\theta d_{2}},&X_{141}=c_{4}e^{i\theta e_{1}},&X_{142}=c_{4}e^{i\theta e_{4}}, &X_{143}=c_{4}e^{i\theta f_{1}},    &  X_{144}=c_{4}e^{i\theta f_{3}},  \\
X_{145}=d_{3}e^{i\theta a_{2}},&X_{146}=d_{3}e^{i\theta a_{3}},&X_{147}=d_{3}e^{i\theta a_{4}},&X_{148}=d_{3}e^{i\theta d_{1}}, &X_{149}=d_{3}e^{i\theta d_{2}},    &  X_{150}=d_{3}e^{i\theta e_{1}},  \\
X_{151}=d_{3}e^{i\theta e_{4}},&X_{152}=d_{3}e^{i\theta f_{1}},&X_{153}=d_{3}e^{i\theta f_{3}},&X_{154}=d_{4}e^{i\theta a_{2}}, &X_{155}=d_{4}e^{i\theta a_{3}},   &  X_{156}=d_{4}e^{i\theta a_{4}},  \\
X_{157}=d_{4}e^{i\theta d_{1}},&X_{158}=d_{4}e^{i\theta d_{2}},&X_{159}=d_{4}e^{i\theta e_{1}},&X_{160}=d_{4}e^{i\theta e_{4}}, &X_{161}=d_{4}e^{i\theta f_{1}},   &  X_{162}=d_{4}e^{i\theta f_{3}}, \\
X_{163}=e_{2}e^{i\theta a_{2}},&X_{164}=e_{2}e^{i\theta a_{3}},&X_{165}=e_{2}e^{i\theta a_{4}},&X_{166}=e_{2}e^{i\theta d_{1}}, &X_{167}=e_{2}e^{i\theta d_{2}},   &  X_{168}=e_{2}e^{i\theta e_{1}},
\end{array}
\end{equation}
\begin{equation}\nonumber
\begin{array}{cccccc}
X_{169}=e_{2}e^{i\theta e_{4}},&X_{170}=e_{2}e^{i\theta f_{1}},&X_{171}=e_{2}e^{i\theta f_{3}},&X_{172}=e_{3}e^{i\theta a_{2}}, &X_{173}=e_{3}e^{i\theta a_{3}},   &  X_{174}=e_{3}e^{i\theta a_{4}},\\
X_{175}=e_{3}e^{i\theta d_{1}},&X_{176}=e_{3}e^{i\theta d_{2}},&X_{177}=e_{3}e^{i\theta e_{1}},&X_{178}=e_{3}e^{i\theta e_{4}}, &X_{179}=e_{3}e^{i\theta f_{1}},   &  X_{180}=e_{3}e^{i\theta f_{3}}, \\
X_{181}=f_{2}e^{i\theta a_{2}},&X_{182}=f_{2}e^{i\theta a_{3}},&X_{183}=f_{2}e^{i\theta a_{4}},&X_{184}=f_{2}e^{i\theta d_{1}}, &X_{185}=f_{2}e^{i\theta d_{2}},   &  X_{186}=f_{2}e^{i\theta e_{1}},\\
X_{187}=f_{2}e^{i\theta e_{4}},&X_{188}=f_{2}e^{i\theta f_{1}},&X_{189}=f_{2}e^{i\theta f_{3}},&X_{190}=f_{4}e^{i\theta a_{2}}, &X_{191}=f_{4}e^{i\theta a_{3}},   &  X_{192}=f_{4}e^{i\theta a_{4}},\\
X_{193}=f_{4}e^{i\theta d_{1}},&X_{194}=f_{4}e^{i\theta d_{2}},&X_{195}=f_{4}e^{i\theta e_{1}},&X_{196}=f_{4}e^{i\theta e_{4}}, &X_{197}=f_{4}e^{i\theta f_{1}},   &  X_{198}=f_{4}e^{i\theta f_{3}}.
\end{array}
\end{equation}

We note that if the identity element is used in the construction of GA, no novel symmetries would be emergent beyond the generating group itself. Hence, we just consider the GA elements listed above.

\subsection{Classification of the elements of C[$S_{4}$]}
The symmetry expressed by a C[$S_{4}$] element is determined by the following relation
\begin{equation}\label{eq:10}
X^{+}M^{+}MX=M^{+}M,
\end{equation}
where $M$ is the mass matrix of charged leptons or Dirac neutrinos. On the basis of this constraint, the mass matrix may
show special symmetries  which are apparently different from $X$ itself. For example, $M$ may pose the following property:
\begin{equation}\label{eq:11}
g^{+}M^{+}Mg=M^{+}M,~~~CP^{+}(M^{+}M)CP=(M^{+}M)^{*},
\end{equation}
where $g\neq X$, $CP$ denotes a generalised CP transformation. Thus, the Eq.\ref{eq:10} determines  the equivalent symmetry of $X$.
For instance, the element $X_{29}$ corresponds to the equivalent symmetry $Z_{2}^{C_{magic}}\times CP$\cite{Rong:2019qjh}. Here $C_{magic}$ is the magic matrix, namely
\begin{equation}
\label{eq:12}
C_{magic}=\frac{1}{3}\left(
            \begin{array}{ccc}
              1 & -2 & -2 \\
              -2 & 1 & -2 \\
              -2 & -2 & 1 \\
            \end{array}
          \right), ~~~C_{magic}^{2}=I,
\end{equation}
$CP$ is the 2-3 permutation matrix $d_{1}$ given in Eq.\ref{eq:8}. $Z_{2}^{C_{magic}}$ denotes the  group $Z_{2}$ generated by the generator  $C_{magic}$.
 According to the equivalent symmetries of the GA, we perform a classification of the elements. There are 7 types listed as follows:
 \begin{equation} \label{eq:13}
  A^{(6)}=\{X_{1},~X_{2},~X_{9},~X_{10},~X_{17},~X_{18}\},
 \end{equation}
  \begin{equation} \label{eq:14}
  \begin{array}{c}
    B_{1}^{(18)}= \{X_{3},  X_{4}, X_{15},  X_{16},  X_{21},  X_{22},  X_{25}, X_{28},  X_{33}, \\
   ~~~~~~~~ X_{36},  X_{43},  X_{46},  X_{51},  X_{54},  X_{58},  X_{64},  X_{66},  X_{72}\},
  \end{array}
 \end{equation}
  \begin{equation} \label{eq:15}
  \begin{array}{c}
    B_{2}^{(24)}= \{X_{5},  X_{6}, X_{7},  X_{8},  X_{11},  X_{12},  X_{13}, X_{14},  X_{19}, X_{20},  X_{23}, X_{24}, \\
 ~~~~~~~~X_{148},  X_{149}, X_{157},  X_{158},  X_{168},  X_{169},  X_{177}, X_{178},  X_{188}, X_{189},  X_{197}, X_{198}\},
  \end{array}
 \end{equation}
  \begin{equation} \label{eq:16}
  \begin{array}{c}
    B_{3}^{(30)}= \{X_{26},  X_{27}, X_{34},  X_{35},  X_{41},  X_{42},  X_{49}, X_{50},  X_{57}, X_{59},  X_{65}, X_{67}, X_{145},  X_{146}, X_{147}, \\
 ~~~~~~~~X_{154},  X_{155}, X_{156},  X_{163},  X_{164},  X_{165},  X_{172}, X_{173},  X_{174}, X_{181},  X_{182}, X_{183},X_{190},  X_{191}, X_{192}\},
  \end{array}
 \end{equation}
   \begin{equation} \label{eq:17}
  \begin{array}{c}
    C_{1}^{(48)}= \{X_{29},  X_{30}, X_{31},  X_{32},  X_{37},  X_{38},  X_{39}, X_{40},  X_{44}, X_{45},  X_{47}, X_{48}, X_{52},  X_{53}, X_{55},  X_{56},\\
 ~~~X_{60},  X_{61}, X_{62},  X_{63},  X_{68},  X_{69},  X_{70}, X_{71},  X_{76}, X_{78},  X_{80}, X_{86},X_{88},  X_{89}, X_{95}, X_{96}\\
~~~X_{99},  X_{103}, X_{106},  X_{108},  X_{112},  X_{114},  X_{116}, X_{122},  X_{123}, X_{126},  X_{130}, X_{133},X_{135},  X_{140}, X_{142}, X_{143}\},
  \end{array}
 \end{equation}
    \begin{equation} \label{eq:18}
  \begin{array}{c}
    C_{2}^{(48)}= \{X_{77},  X_{79}, X_{81},  X_{85},  X_{87},  X_{90},  X_{94}, X_{97},  X_{98}, X_{104},  X_{105}, X_{107}, X_{113},  X_{115}, X_{117},  X_{121},\\
 ~~~X_{124},  X_{125}, X_{131},  X_{132},  X_{134},  X_{139},  X_{141}, X_{144},  X_{150}, X_{151},  X_{152}, X_{153},X_{159},  X_{160}, X_{161}, X_{162}\\
~~~X_{166},  X_{167}, X_{170},  X_{171},  X_{175},  X_{176},  X_{179}, X_{180},  X_{184}, X_{185},  X_{186}, X_{187},X_{193},  X_{194}, X_{195}, X_{196}\},
  \end{array}
 \end{equation}
 \begin{equation} \label{eq:19}
  \begin{array}{c}
   D^{(24)}= \{X_{73},  X_{74}, X_{75},  X_{82},  X_{83},  X_{84},  X_{91}, X_{92},  X_{93}, X_{100},  X_{101}, X_{102}, \\
 ~~~X_{109},  X_{110}, X_{111},  X_{118},  X_{119},  X_{120},  X_{127}, X_{128},  X_{129}, X_{136},  X_{137}, X_{138}\}.
  \end{array}
 \end{equation}
 Here the superscript $(j)$ denotes that the number of the elements in the type is $j$.
  The specific symmetries expressed by the representative in every type are shown in Tab. \ref{Tab:1}.
   \begin{table}
	\caption{\label{Tab:1} Symmetries expressed by the representative elements.}
	\centering
	\begin{tabular}{ccc}
		\hline
Type&~~~~~~Representative~~~~~~&~~~~~~Equivalent symmetry~~~~~~\\
		\hline
$A^{(6)}$&~~~~~~$X_{1}=\left(
                  \begin{array}{ccc}
                    e^{-i\theta} & 0 & 0 \\
                    0& -e^{i\theta} & 0 \\
                    0 & 0 & -e^{-i\theta} \\
                  \end{array}
                \right)$~~~~~&~~~~~~~~$U(1)\times U(1)\times U(1)$~~~~~~\\
$B_{1}^{(18)}$&~~~~~~~$X_{3}=\left(
                  \begin{array}{ccc}
                    e^{i\theta} & 0 & 0 \\
                    0&-\cos\theta & -i\sin\theta \\
                    0 & -i\sin\theta & -\cos\theta  \\
                  \end{array}
                \right)$~~~~~~&~~~~~~~~~$U(1)\times Z_{2}^{d_{1}}$~~~~~~\\
$B_{2}^{(24)}$&~~~~~$X_{5}=\left(
                  \begin{array}{ccc}
                    \cos\theta & i\sin\theta & 0 \\
                    -i\sin\theta&-\cos\theta & 0\\
                    0 & 0 & -e^{i\theta}   \\
                  \end{array}
                \right)$~~~~~~&~~~~~~~~$U(1)$~~~~~~\\
$B_{3}^{(30)}$&~~~~~~$X_{26}=\left(
                  \begin{array}{ccc}
                   e^{-i\theta}  & 0 & 0 \\
                    0&0 & e^{-i\theta} \\
                    0 & e^{i\theta}  & 0   \\
                  \end{array}
                \right)$~~~~~~&~~~~~~~~$U(1)\times Z_{2}^{g}\times CP$, ~~$g=\left(
                  \begin{array}{ccc}
                   1 & 0 & 0 \\
                    0&0 & e^{-i\theta} \\
                    0 & e^{i\theta}  & 0   \\
                  \end{array}
                \right)$, $CP=d_{1}$~~~~\\
$C_{1}^{(48)}$&~~~~~~$X_{29}=\left(
                  \begin{array}{ccc}
                   \cos\theta   & i\sin\theta & 0 \\
                    0&0 & e^{i\theta} \\
                    i\sin\theta & \cos\theta & 0   \\
                  \end{array}
                \right)$~~~~~~&~~~~~~~$Z_{2}^{C_{magic}}\times CP$, $CP=d_{1}$~~~~~\\
$C_{2}^{(48)}$&~~~~~~$X_{77}=\left(
                  \begin{array}{ccc}
                   0   & -i\sin\theta &  \cos\theta  \\
                    e^{i\theta}&0 & 0 \\
                   0 & \cos\theta & i\sin\theta  \\
                  \end{array}
                \right)$~~~~~~&~~~~~$CP$,~~~$CP=e_{1}$~~~~~~\\
$D^{(24)}$&~~~~~~$X_{73}=\left(
                  \begin{array}{ccc}
                   0   & 0&  e^{-i\theta}  \\
                    e^{i\theta}&0 & 0 \\
                   0 &  e^{-i\theta} &0 \\
                  \end{array}
                \right)$ &$\begin{array}{c}
                            Z_{2}^{g}\times CP_{1}\times CP_{2},  \\
                            g=diag(e^{\frac{2i\theta}{3}}, ~e^{\frac{-2i\theta}{3}},~ 1)C_{magic}diag(e^{\frac{-2i\theta}{3}}, ~e^{\frac{2i\theta}{3}},~ 1), \\
                             CP_{1}=e_{1}, ~~CP_{2}=d_{1}diag(e^{-2i\theta}, ~1,~ 1)
                          \end{array}$             \\
		\hline
	\end{tabular}
\end{table}

Let us stress 3 noticeable  observations from Tab. \ref{Tab:1}.\\
(1). On the basis of the constraint Eq.\ref{eq:10}, the equivalent symmetry of a GA element can be continuous (e.g., in the case of Type $A^{(6)}$, $B_{2}^{(24)}$), discrete (e.g., in the case of Type $C_{1}^{(48)}$), and hybrid (e.g., in the case of Type $B_{1}^{(18)}$). Here the basic continuous symmetry is expressed by the 1-dimensional unitary group $U(1)$ which is derived from the diagonal or block-diagonal form of the GA elements.\\
(2). The equivalent symmetry $Z_{2}\times CP$ only takes place in the Type $C_{1}^{(48)}$ and Type $D^{(24)}$. In the former type, a GA element $X$ leads to only one generalised CP symmetry, while in the latter case $X$ brings two CP symmetries. \\
(3). The generalized CP symmetry can appear without the company of a residual flavor symmetry. In the case of Type $C_{2}^{(48)}$, a GA element only generates a CP symmetry without a corresponding $Z_{2}$ symmetry.

\subsection{Some phenomenological discussions }
By now, the origin of leptonic mixing remains a riddle. As theoretically plausible assumptions,  some flavor symmetries and generalised symmetries are proposed to interpret the global fit data of the neutrino oscillation parameters. At the phenomenological level, possible symmetries combinations of the leptonic mass matrices are surveyed\cite{Hernandez:2012ra,Holthausen:2012wt,Ge:2011qn,Grimus:2013rw,Lavoura:2014kwa,Fonseca:2014koa}, e.g. $(Z_{2e}\times CP_{e}, Z_{2\nu}\times CP_{\nu})$\cite{Rong:2016cpk}, where the subscripts $e$ and $\nu$ denote the charged leptons and neutrinos sector respectively. In fact, the novel symmetry combination $(X_{e}, X_{\nu})$ from $C[S_{4}]$ is also viable. To be more specific, we assume the constraints
$X^{+}_{e, \nu}M^{+}_{e, \nu}M_{e, \nu}X_{e, \nu}=M^{+}_{e,\nu}M_{e, \nu}$, and derive the leptonic mixing matrix $U_{PMNS}=U_{e}^{+}U_{\nu}$ from the following relations
\begin{equation}\label{eq:20}
U_{e}^{+}X_{e}U_{e}=diag(e^{i\alpha_{1}},~e^{i\alpha_{2}},~~e^{i\alpha_{3}}),~ U_{\nu}^{+}X_{\nu}U_{\nu}=diag(e^{i\beta_{1}},~e^{i\beta_{2}},~~e^{i\beta_{3}}).
\end{equation}
As an example, the symmetry pair $(X_{e}=X_{1}, X_{\nu}=X_{29})$ gives rise to the TM$_{2}$ mixing pattern, which denotes that the second column of the leptonic mixing matrix is of the form
\begin{equation}\label{eq:21}
\begin{array}{ccc}
  U_{(2)}=(\frac{1}{\sqrt{3}} & \frac{1}{\sqrt{3}} & \frac{1}{\sqrt{3}} )^{T}.
\end{array}
\end{equation}
This pattern is near the margin of the $3\sigma$ allowed ranges of the mixing parameters.
The combination  $(X_{e}=X_{74}, X_{\nu}=X_{77})$ leads to the mixing pattern which can accommodate the global fit data at $1\sigma$ level around the best fit values in the case of normal mass ordering\cite{Esteban:2020cvm}. Considering the inverted mass ordering, the favored CP phase $\delta_{CP}\sim\frac{-\pi}{2}$ could be obtained in the mixing pattern based on the combination $(X_{e}=X_{5}, X_{\nu}=X_{29})$.
Therefore, the generalised symmetry expressed by a GA element can match the reported global fit data of neutrino oscillations.

\section{Conclusions}
Besides discrete flavor groups, a novel mathematical object called GA can also be used to express symmetries of leptonic mass matrices.
A GA element describes a continuous symmetry generated by discrete-group elements. We classified the nontrivial elements of the 2-dimensional GA called $C[S_{4}]$ based on the group $S_{4}$. It is found that the equivalent symmetries derived from  a $C[S_{4}]$ element can be continuous, discrete, and hybrid. The landscape of the symmetries from $C[S_{4}]$ includes $U(1)$, $U(1)\times U(1)\times U(1)$, $U(1)\times Z_{2}$, $U(1)\times Z_{2}\times CP$, $Z_{2}\times CP$, $CP$, and $Z_{2}\times CP_{1}\times CP_{2}$.
Correspondingly, under the constraint of the combination of the symmetries from the GA, the leptonic mass matrices can give rise to the mixing matrix accommodating the global
fit data of neutrino oscillations.

\vspace{0.08cm}

\acknowledgments
We are grateful to Min Wen for his help in the preparation of the work.
This work is supported by the National Natural Science Foundation of China under grant No. 12065007.

\bibliography{refs}

\begin{thebibliography}{38}%
\makeatletter
\providecommand \@ifxundefined [1]{%
 \@ifx{#1\undefined}
}%
\providecommand \@ifnum [1]{%
 \ifnum #1\expandafter \@firstoftwo
 \else \expandafter \@secondoftwo
 \fi
}%
\providecommand \@ifx [1]{%
 \ifx #1\expandafter \@firstoftwo
 \else \expandafter \@secondoftwo
 \fi
}%
\providecommand \natexlab [1]{#1}%
\providecommand \enquote  [1]{``#1''}%
\providecommand \bibnamefont  [1]{#1}%
\providecommand \bibfnamefont [1]{#1}%
\providecommand \citenamefont [1]{#1}%
\providecommand \href@noop [0]{\@secondoftwo}%
\providecommand \href [0]{\begingroup \@sanitize@url \@href}%
\providecommand \@href[1]{\@@startlink{#1}\@@href}%
\providecommand \@@href[1]{\endgroup#1\@@endlink}%
\providecommand \@sanitize@url [0]{\catcode `\\12\catcode `\$12\catcode
  `\&12\catcode `\#12\catcode `\^12\catcode `\_12\catcode `\%12\relax}%
\providecommand \@@startlink[1]{}%
\providecommand \@@endlink[0]{}%
\providecommand \url  [0]{\begingroup\@sanitize@url \@url }%
\providecommand \@url [1]{\endgroup\@href {#1}{\urlprefix }}%
\providecommand \urlprefix  [0]{URL }%
\providecommand \Eprint [0]{\href }%
\providecommand \doibase [0]{http://dx.doi.org/}%
\providecommand \selectlanguage [0]{\@gobble}%
\providecommand \bibinfo  [0]{\@secondoftwo}%
\providecommand \bibfield  [0]{\@secondoftwo}%
\providecommand \translation [1]{[#1]}%
\providecommand \BibitemOpen [0]{}%
\providecommand \bibitemStop [0]{}%
\providecommand \bibitemNoStop [0]{.\EOS\space}%
\providecommand \EOS [0]{\spacefactor3000\relax}%
\providecommand \BibitemShut  [1]{\csname bibitem#1\endcsname}%
\let\auto@bib@innerbib\@empty
\bibitem [{\citenamefont {Yanagida}(1980)}]{Yanagida:1980xy}%
  \BibitemOpen
  \bibfield  {author} {\bibinfo {author} {\bibfnamefont {T.}~\bibnamefont
  {Yanagida}},\ }\href {\doibase 10.1143/PTP.64.1103} {\bibfield  {journal}
  {\bibinfo  {journal} {Prog. Theor. Phys.}\ }\textbf {\bibinfo {volume}
  {64}},\ \bibinfo {pages} {1103} (\bibinfo {year} {1980})}\BibitemShut
  {NoStop}%
\bibitem [{\citenamefont {Ohlsson}\ and\ \citenamefont
  {Seidl}(2002)}]{Ohlsson:2002rb}%
  \BibitemOpen
  \bibfield  {author} {\bibinfo {author} {\bibfnamefont {T.}~\bibnamefont
  {Ohlsson}}\ and\ \bibinfo {author} {\bibfnamefont {G.}~\bibnamefont
  {Seidl}},\ }\href {\doibase 10.1016/S0550-3213(02)00689-2} {\bibfield
  {journal} {\bibinfo  {journal} {Nucl. Phys. B}\ }\textbf {\bibinfo {volume}
  {643}},\ \bibinfo {pages} {247} (\bibinfo {year} {2002})},\ \Eprint
  {http://arxiv.org/abs/hep-ph/0206087} {arXiv:hep-ph/0206087} \BibitemShut
  {NoStop}%
\bibitem [{\citenamefont {Babu}\ \emph {et~al.}(2003)\citenamefont {Babu},
  \citenamefont {Ma},\ and\ \citenamefont {Valle}}]{Babu:2002dz}%
  \BibitemOpen
  \bibfield  {author} {\bibinfo {author} {\bibfnamefont {K.~S.}\ \bibnamefont
  {Babu}}, \bibinfo {author} {\bibfnamefont {E.}~\bibnamefont {Ma}}, \ and\
  \bibinfo {author} {\bibfnamefont {J.~W.~F.}\ \bibnamefont {Valle}},\ }\href
  {\doibase 10.1016/S0370-2693(02)03153-2} {\bibfield  {journal} {\bibinfo
  {journal} {Phys. Lett. B}\ }\textbf {\bibinfo {volume} {552}},\ \bibinfo
  {pages} {207} (\bibinfo {year} {2003})},\ \Eprint
  {http://arxiv.org/abs/hep-ph/0206292} {arXiv:hep-ph/0206292} \BibitemShut
  {NoStop}%
\bibitem [{\citenamefont {Ma}(2004)}]{Ma:2004zv}%
  \BibitemOpen
  \bibfield  {author} {\bibinfo {author} {\bibfnamefont {E.}~\bibnamefont
  {Ma}},\ }\href {\doibase 10.1103/PhysRevD.70.031901} {\bibfield  {journal}
  {\bibinfo  {journal} {Phys. Rev. D}\ }\textbf {\bibinfo {volume} {70}},\
  \bibinfo {pages} {031901} (\bibinfo {year} {2004})},\ \Eprint
  {http://arxiv.org/abs/hep-ph/0404199} {arXiv:hep-ph/0404199} \BibitemShut
  {NoStop}%
\bibitem [{\citenamefont {Grimus}\ and\ \citenamefont
  {Lavoura}(2006)}]{Grimus:2005rf}%
  \BibitemOpen
  \bibfield  {author} {\bibinfo {author} {\bibfnamefont {W.}~\bibnamefont
  {Grimus}}\ and\ \bibinfo {author} {\bibfnamefont {L.}~\bibnamefont
  {Lavoura}},\ }\href {\doibase 10.1088/1126-6708/2006/01/018} {\bibfield
  {journal} {\bibinfo  {journal} {JHEP}\ }\textbf {\bibinfo {volume} {01}},\
  \bibinfo {pages} {018} (\bibinfo {year} {2006})},\ \Eprint
  {http://arxiv.org/abs/hep-ph/0509239} {arXiv:hep-ph/0509239} \BibitemShut
  {NoStop}%
\bibitem [{\citenamefont {Lam}(2006)}]{Lam:2006wm}%
  \BibitemOpen
  \bibfield  {author} {\bibinfo {author} {\bibfnamefont {C.~S.}\ \bibnamefont
  {Lam}},\ }\href {\doibase 10.1103/PhysRevD.74.113004} {\bibfield  {journal}
  {\bibinfo  {journal} {Phys. Rev. D}\ }\textbf {\bibinfo {volume} {74}},\
  \bibinfo {pages} {113004} (\bibinfo {year} {2006})},\ \Eprint
  {http://arxiv.org/abs/hep-ph/0611017} {arXiv:hep-ph/0611017} \BibitemShut
  {NoStop}%
\bibitem [{\citenamefont {Xing}\ \emph {et~al.}(2006)\citenamefont {Xing},
  \citenamefont {Zhang},\ and\ \citenamefont {Zhou}}]{Xing:2006xa}%
  \BibitemOpen
  \bibfield  {author} {\bibinfo {author} {\bibfnamefont {Z.-z.}\ \bibnamefont
  {Xing}}, \bibinfo {author} {\bibfnamefont {H.}~\bibnamefont {Zhang}}, \ and\
  \bibinfo {author} {\bibfnamefont {S.}~\bibnamefont {Zhou}},\ }\href {\doibase
  10.1016/j.physletb.2006.08.045} {\bibfield  {journal} {\bibinfo  {journal}
  {Phys. Lett. B}\ }\textbf {\bibinfo {volume} {641}},\ \bibinfo {pages} {189}
  (\bibinfo {year} {2006})},\ \Eprint {http://arxiv.org/abs/hep-ph/0607091}
  {arXiv:hep-ph/0607091} \BibitemShut {NoStop}%
\bibitem [{\citenamefont {He}\ \emph {et~al.}(2006)\citenamefont {He},
  \citenamefont {Keum},\ and\ \citenamefont {Volkas}}]{He:2006dk}%
  \BibitemOpen
  \bibfield  {author} {\bibinfo {author} {\bibfnamefont {X.-G.}\ \bibnamefont
  {He}}, \bibinfo {author} {\bibfnamefont {Y.-Y.}\ \bibnamefont {Keum}}, \ and\
  \bibinfo {author} {\bibfnamefont {R.~R.}\ \bibnamefont {Volkas}},\ }\href
  {\doibase 10.1088/1126-6708/2006/04/039} {\bibfield  {journal} {\bibinfo
  {journal} {JHEP}\ }\textbf {\bibinfo {volume} {04}},\ \bibinfo {pages} {039}
  (\bibinfo {year} {2006})},\ \Eprint {http://arxiv.org/abs/hep-ph/0601001}
  {arXiv:hep-ph/0601001} \BibitemShut {NoStop}%
\bibitem [{\citenamefont {Altarelli}\ \emph {et~al.}(2009)\citenamefont
  {Altarelli}, \citenamefont {Feruglio},\ and\ \citenamefont
  {Merlo}}]{Altarelli:2009gn}%
  \BibitemOpen
  \bibfield  {author} {\bibinfo {author} {\bibfnamefont {G.}~\bibnamefont
  {Altarelli}}, \bibinfo {author} {\bibfnamefont {F.}~\bibnamefont {Feruglio}},
  \ and\ \bibinfo {author} {\bibfnamefont {L.}~\bibnamefont {Merlo}},\ }\href
  {\doibase 10.1088/1126-6708/2009/05/020} {\bibfield  {journal} {\bibinfo
  {journal} {JHEP}\ }\textbf {\bibinfo {volume} {05}},\ \bibinfo {pages} {020}
  (\bibinfo {year} {2009})},\ \Eprint {http://arxiv.org/abs/0903.1940}
  {arXiv:0903.1940 [hep-ph]} \BibitemShut {NoStop}%
\bibitem [{\citenamefont {Chen}\ and\ \citenamefont
  {King}(2009)}]{Chen:2009um}%
  \BibitemOpen
  \bibfield  {author} {\bibinfo {author} {\bibfnamefont {M.-C.}\ \bibnamefont
  {Chen}}\ and\ \bibinfo {author} {\bibfnamefont {S.~F.}\ \bibnamefont
  {King}},\ }\href {\doibase 10.1088/1126-6708/2009/06/072} {\bibfield
  {journal} {\bibinfo  {journal} {JHEP}\ }\textbf {\bibinfo {volume} {06}},\
  \bibinfo {pages} {072} (\bibinfo {year} {2009})},\ \Eprint
  {http://arxiv.org/abs/0903.0125} {arXiv:0903.0125 [hep-ph]} \BibitemShut
  {NoStop}%
\bibitem [{\citenamefont {Antusch}\ and\ \citenamefont
  {Maurer}(2011)}]{Antusch:2011qg}%
  \BibitemOpen
  \bibfield  {author} {\bibinfo {author} {\bibfnamefont {S.}~\bibnamefont
  {Antusch}}\ and\ \bibinfo {author} {\bibfnamefont {V.}~\bibnamefont
  {Maurer}},\ }\href {\doibase 10.1103/PhysRevD.84.117301} {\bibfield
  {journal} {\bibinfo  {journal} {Phys. Rev. D}\ }\textbf {\bibinfo {volume}
  {84}},\ \bibinfo {pages} {117301} (\bibinfo {year} {2011})},\ \Eprint
  {http://arxiv.org/abs/1107.3728} {arXiv:1107.3728 [hep-ph]} \BibitemShut
  {NoStop}%
\bibitem [{\citenamefont {Hernandez}\ and\ \citenamefont
  {Smirnov}(2012)}]{Hernandez:2012ra}%
  \BibitemOpen
  \bibfield  {author} {\bibinfo {author} {\bibfnamefont {D.}~\bibnamefont
  {Hernandez}}\ and\ \bibinfo {author} {\bibfnamefont {A.~Y.}\ \bibnamefont
  {Smirnov}},\ }\href {\doibase 10.1103/PhysRevD.86.053014} {\bibfield
  {journal} {\bibinfo  {journal} {Phys. Rev. D}\ }\textbf {\bibinfo {volume}
  {86}},\ \bibinfo {pages} {053014} (\bibinfo {year} {2012})},\ \Eprint
  {http://arxiv.org/abs/1204.0445} {arXiv:1204.0445 [hep-ph]} \BibitemShut
  {NoStop}%
\bibitem [{\citenamefont {Feruglio}(2019)}]{Feruglio:2017spp}%
  \BibitemOpen
  \bibfield  {author} {\bibinfo {author} {\bibfnamefont {F.}~\bibnamefont
  {Feruglio}},\ }\href@noop {} {\  (\bibinfo {year} {2019})},\ \Eprint
  {http://arxiv.org/abs/1706.08749} {arXiv:1706.08749 [hep-ph]} \BibitemShut
  {NoStop}%
\bibitem [{\citenamefont {Altarelli}\ and\ \citenamefont
  {Feruglio}(2010)}]{Altarelli:2010gt}%
  \BibitemOpen
  \bibfield  {author} {\bibinfo {author} {\bibfnamefont {G.}~\bibnamefont
  {Altarelli}}\ and\ \bibinfo {author} {\bibfnamefont {F.}~\bibnamefont
  {Feruglio}},\ }\href {\doibase 10.1103/RevModPhys.82.2701} {\bibfield
  {journal} {\bibinfo  {journal} {Rev. Mod. Phys.}\ }\textbf {\bibinfo {volume}
  {82}},\ \bibinfo {pages} {2701} (\bibinfo {year} {2010})},\ \Eprint
  {http://arxiv.org/abs/1002.0211} {arXiv:1002.0211 [hep-ph]} \BibitemShut
  {NoStop}%
\bibitem [{\citenamefont {King}\ and\ \citenamefont
  {Luhn}(2013)}]{King:2013eh}%
  \BibitemOpen
  \bibfield  {author} {\bibinfo {author} {\bibfnamefont {S.~F.}\ \bibnamefont
  {King}}\ and\ \bibinfo {author} {\bibfnamefont {C.}~\bibnamefont {Luhn}},\
  }\href {\doibase 10.1088/0034-4885/76/5/056201} {\bibfield  {journal}
  {\bibinfo  {journal} {Rept. Prog. Phys.}\ }\textbf {\bibinfo {volume} {76}},\
  \bibinfo {pages} {056201} (\bibinfo {year} {2013})},\ \Eprint
  {http://arxiv.org/abs/1301.1340} {arXiv:1301.1340 [hep-ph]} \BibitemShut
  {NoStop}%
\bibitem [{\citenamefont {Xing}\ and\ \citenamefont
  {Zhao}(2016)}]{Xing:2015fdg}%
  \BibitemOpen
  \bibfield  {author} {\bibinfo {author} {\bibfnamefont {Z.-z.}\ \bibnamefont
  {Xing}}\ and\ \bibinfo {author} {\bibfnamefont {Z.-h.}\ \bibnamefont
  {Zhao}},\ }\href {\doibase 10.1088/0034-4885/79/7/076201} {\bibfield
  {journal} {\bibinfo  {journal} {Rept. Prog. Phys.}\ }\textbf {\bibinfo
  {volume} {79}},\ \bibinfo {pages} {076201} (\bibinfo {year} {2016})},\
  \Eprint {http://arxiv.org/abs/1512.04207} {arXiv:1512.04207 [hep-ph]}
  \BibitemShut {NoStop}%
\bibitem [{\citenamefont {Feruglio}\ and\ \citenamefont
  {Romanino}(2021)}]{Feruglio:2019ybq}%
  \BibitemOpen
  \bibfield  {author} {\bibinfo {author} {\bibfnamefont {F.}~\bibnamefont
  {Feruglio}}\ and\ \bibinfo {author} {\bibfnamefont {A.}~\bibnamefont
  {Romanino}},\ }\href {\doibase 10.1103/RevModPhys.93.015007} {\bibfield
  {journal} {\bibinfo  {journal} {Rev. Mod. Phys.}\ }\textbf {\bibinfo {volume}
  {93}},\ \bibinfo {pages} {015007} (\bibinfo {year} {2021})},\ \Eprint
  {http://arxiv.org/abs/1912.06028} {arXiv:1912.06028 [hep-ph]} \BibitemShut
  {NoStop}%
\bibitem [{\citenamefont {Xing}(2020)}]{Xing:2020ijf}%
  \BibitemOpen
  \bibfield  {author} {\bibinfo {author} {\bibfnamefont {Z.-z.}\ \bibnamefont
  {Xing}},\ }\href {\doibase 10.1016/j.physrep.2020.02.001} {\bibfield
  {journal} {\bibinfo  {journal} {Phys. Rept.}\ }\textbf {\bibinfo {volume}
  {854}},\ \bibinfo {pages} {1} (\bibinfo {year} {2020})},\ \Eprint
  {http://arxiv.org/abs/1909.09610} {arXiv:1909.09610 [hep-ph]} \BibitemShut
  {NoStop}%
\bibitem [{\citenamefont {Ding}\ and\ \citenamefont
  {King}(2024)}]{Ding:2023htn}%
  \BibitemOpen
  \bibfield  {author} {\bibinfo {author} {\bibfnamefont {G.-J.}\ \bibnamefont
  {Ding}}\ and\ \bibinfo {author} {\bibfnamefont {S.~F.}\ \bibnamefont
  {King}},\ }\href {\doibase 10.1088/1361-6633/ad52a3} {\bibfield  {journal}
  {\bibinfo  {journal} {Rept. Prog. Phys.}\ }\textbf {\bibinfo {volume} {87}},\
  \bibinfo {pages} {084201} (\bibinfo {year} {2024})},\ \Eprint
  {http://arxiv.org/abs/2311.09282} {arXiv:2311.09282 [hep-ph]} \BibitemShut
  {NoStop}%
\bibitem [{\citenamefont {Ding}\ and\ \citenamefont
  {Valle}(2024)}]{Ding:2024ozt}%
  \BibitemOpen
  \bibfield  {author} {\bibinfo {author} {\bibfnamefont {G.-J.}\ \bibnamefont
  {Ding}}\ and\ \bibinfo {author} {\bibfnamefont {J.~W.~F.}\ \bibnamefont
  {Valle}},\ }\href@noop {} {\  (\bibinfo {year} {2024})},\ \Eprint
  {http://arxiv.org/abs/2402.16963} {arXiv:2402.16963 [hep-ph]} \BibitemShut
  {NoStop}%
\bibitem [{\citenamefont {Rong}(2017{\natexlab{a}})}]{Rong:2017rel}%
  \BibitemOpen
  \bibfield  {author} {\bibinfo {author} {\bibfnamefont {S.-j.}\ \bibnamefont
  {Rong}},\ }\href@noop {} {\  (\bibinfo {year} {2017}{\natexlab{a}})},\
  \Eprint {http://arxiv.org/abs/1703.09981} {arXiv:1703.09981 [hep-ph]}
  \BibitemShut {NoStop}%
\bibitem [{\citenamefont {Rong}(2020)}]{Rong:2019qjh}%
  \BibitemOpen
  \bibfield  {author} {\bibinfo {author} {\bibfnamefont {S.-J.}\ \bibnamefont
  {Rong}},\ }\href {\doibase 10.1155/2020/3967605} {\bibfield  {journal}
  {\bibinfo  {journal} {Adv. High Energy Phys.}\ }\textbf {\bibinfo {volume}
  {2020}},\ \bibinfo {pages} {3967605} (\bibinfo {year} {2020})},\ \Eprint
  {http://arxiv.org/abs/1906.09157} {arXiv:1906.09157 [hep-ph]} \BibitemShut
  {NoStop}%
\bibitem [{\citenamefont {Grimus}\ and\ \citenamefont
  {Lavoura}(2004)}]{Grimus:2003yn}%
  \BibitemOpen
  \bibfield  {author} {\bibinfo {author} {\bibfnamefont {W.}~\bibnamefont
  {Grimus}}\ and\ \bibinfo {author} {\bibfnamefont {L.}~\bibnamefont
  {Lavoura}},\ }\href {\doibase 10.1016/j.physletb.2003.10.075} {\bibfield
  {journal} {\bibinfo  {journal} {Phys. Lett. B}\ }\textbf {\bibinfo {volume}
  {579}},\ \bibinfo {pages} {113} (\bibinfo {year} {2004})},\ \Eprint
  {http://arxiv.org/abs/hep-ph/0305309} {arXiv:hep-ph/0305309} \BibitemShut
  {NoStop}%
\bibitem [{\citenamefont {Feruglio}\ \emph {et~al.}(2013)\citenamefont
  {Feruglio}, \citenamefont {Hagedorn},\ and\ \citenamefont
  {Ziegler}}]{Feruglio:2012cw}%
  \BibitemOpen
  \bibfield  {author} {\bibinfo {author} {\bibfnamefont {F.}~\bibnamefont
  {Feruglio}}, \bibinfo {author} {\bibfnamefont {C.}~\bibnamefont {Hagedorn}},
  \ and\ \bibinfo {author} {\bibfnamefont {R.}~\bibnamefont {Ziegler}},\ }\href
  {\doibase 10.1007/JHEP07(2013)027} {\bibfield  {journal} {\bibinfo  {journal}
  {JHEP}\ }\textbf {\bibinfo {volume} {07}},\ \bibinfo {pages} {027} (\bibinfo
  {year} {2013})},\ \Eprint {http://arxiv.org/abs/1211.5560} {arXiv:1211.5560
  [hep-ph]} \BibitemShut {NoStop}%
\bibitem [{\citenamefont {Feruglio}\ \emph {et~al.}(2014)\citenamefont
  {Feruglio}, \citenamefont {Hagedorn},\ and\ \citenamefont
  {Ziegler}}]{Feruglio:2013hia}%
  \BibitemOpen
  \bibfield  {author} {\bibinfo {author} {\bibfnamefont {F.}~\bibnamefont
  {Feruglio}}, \bibinfo {author} {\bibfnamefont {C.}~\bibnamefont {Hagedorn}},
  \ and\ \bibinfo {author} {\bibfnamefont {R.}~\bibnamefont {Ziegler}},\ }\href
  {\doibase 10.1140/epjc/s10052-014-2753-2} {\bibfield  {journal} {\bibinfo
  {journal} {Eur. Phys. J. C}\ }\textbf {\bibinfo {volume} {74}},\ \bibinfo
  {pages} {2753} (\bibinfo {year} {2014})},\ \Eprint
  {http://arxiv.org/abs/1303.7178} {arXiv:1303.7178 [hep-ph]} \BibitemShut
  {NoStop}%
\bibitem [{\citenamefont {Holthausen}\ \emph
  {et~al.}(2013{\natexlab{a}})\citenamefont {Holthausen}, \citenamefont
  {Lindner},\ and\ \citenamefont {Schmidt}}]{Holthausen:2012dk}%
  \BibitemOpen
  \bibfield  {author} {\bibinfo {author} {\bibfnamefont {M.}~\bibnamefont
  {Holthausen}}, \bibinfo {author} {\bibfnamefont {M.}~\bibnamefont {Lindner}},
  \ and\ \bibinfo {author} {\bibfnamefont {M.~A.}\ \bibnamefont {Schmidt}},\
  }\href {\doibase 10.1007/JHEP04(2013)122} {\bibfield  {journal} {\bibinfo
  {journal} {JHEP}\ }\textbf {\bibinfo {volume} {04}},\ \bibinfo {pages} {122}
  (\bibinfo {year} {2013}{\natexlab{a}})},\ \Eprint
  {http://arxiv.org/abs/1211.6953} {arXiv:1211.6953 [hep-ph]} \BibitemShut
  {NoStop}%
\bibitem [{\citenamefont {Ding}\ \emph {et~al.}(2013)\citenamefont {Ding},
  \citenamefont {King},\ and\ \citenamefont {Stuart}}]{Ding:2013bpa}%
  \BibitemOpen
  \bibfield  {author} {\bibinfo {author} {\bibfnamefont {G.-J.}\ \bibnamefont
  {Ding}}, \bibinfo {author} {\bibfnamefont {S.~F.}\ \bibnamefont {King}}, \
  and\ \bibinfo {author} {\bibfnamefont {A.~J.}\ \bibnamefont {Stuart}},\
  }\href {\doibase 10.1007/JHEP12(2013)006} {\bibfield  {journal} {\bibinfo
  {journal} {JHEP}\ }\textbf {\bibinfo {volume} {12}},\ \bibinfo {pages} {006}
  (\bibinfo {year} {2013})},\ \Eprint {http://arxiv.org/abs/1307.4212}
  {arXiv:1307.4212 [hep-ph]} \BibitemShut {NoStop}%
\bibitem [{\citenamefont {Rong}(2017{\natexlab{b}})}]{Rong:2016cpk}%
  \BibitemOpen
  \bibfield  {author} {\bibinfo {author} {\bibfnamefont {S.-j.}\ \bibnamefont
  {Rong}},\ }\href {\doibase 10.1103/PhysRevD.95.076014} {\bibfield  {journal}
  {\bibinfo  {journal} {Phys. Rev. D}\ }\textbf {\bibinfo {volume} {95}},\
  \bibinfo {pages} {076014} (\bibinfo {year} {2017}{\natexlab{b}})},\ \Eprint
  {http://arxiv.org/abs/1604.08482} {arXiv:1604.08482 [hep-ph]} \BibitemShut
  {NoStop}%
\bibitem [{\citenamefont {Chen}\ \emph {et~al.}(2018)\citenamefont {Chen},
  \citenamefont {Centelles~Chuli\'a}, \citenamefont {Ding}, \citenamefont
  {Srivastava},\ and\ \citenamefont {Valle}}]{Chen:2018lsv}%
  \BibitemOpen
  \bibfield  {author} {\bibinfo {author} {\bibfnamefont {P.}~\bibnamefont
  {Chen}}, \bibinfo {author} {\bibfnamefont {S.}~\bibnamefont
  {Centelles~Chuli\'a}}, \bibinfo {author} {\bibfnamefont {G.-J.}\ \bibnamefont
  {Ding}}, \bibinfo {author} {\bibfnamefont {R.}~\bibnamefont {Srivastava}}, \
  and\ \bibinfo {author} {\bibfnamefont {J.~W.~F.}\ \bibnamefont {Valle}},\
  }\href {\doibase 10.1007/JHEP07(2018)077} {\bibfield  {journal} {\bibinfo
  {journal} {JHEP}\ }\textbf {\bibinfo {volume} {07}},\ \bibinfo {pages} {077}
  (\bibinfo {year} {2018})},\ \Eprint {http://arxiv.org/abs/1802.04275}
  {arXiv:1802.04275 [hep-ph]} \BibitemShut {NoStop}%
\bibitem [{\citenamefont {Novichkov}\ \emph {et~al.}(2019)\citenamefont
  {Novichkov}, \citenamefont {Penedo}, \citenamefont {Petcov},\ and\
  \citenamefont {Titov}}]{Novichkov:2019sqv}%
  \BibitemOpen
  \bibfield  {author} {\bibinfo {author} {\bibfnamefont {P.~P.}\ \bibnamefont
  {Novichkov}}, \bibinfo {author} {\bibfnamefont {J.~T.}\ \bibnamefont
  {Penedo}}, \bibinfo {author} {\bibfnamefont {S.~T.}\ \bibnamefont {Petcov}},
  \ and\ \bibinfo {author} {\bibfnamefont {A.~V.}\ \bibnamefont {Titov}},\
  }\href {\doibase 10.1007/JHEP07(2019)165} {\bibfield  {journal} {\bibinfo
  {journal} {JHEP}\ }\textbf {\bibinfo {volume} {07}},\ \bibinfo {pages} {165}
  (\bibinfo {year} {2019})},\ \Eprint {http://arxiv.org/abs/1905.11970}
  {arXiv:1905.11970 [hep-ph]} \BibitemShut {NoStop}%
\bibitem [{\citenamefont {Nilles}\ \emph {et~al.}(2020)\citenamefont {Nilles},
  \citenamefont {Ramos-S\'anchez},\ and\ \citenamefont
  {Vaudrevange}}]{Nilles:2020nnc}%
  \BibitemOpen
  \bibfield  {author} {\bibinfo {author} {\bibfnamefont {H.~P.}\ \bibnamefont
  {Nilles}}, \bibinfo {author} {\bibfnamefont {S.}~\bibnamefont
  {Ramos-S\'anchez}}, \ and\ \bibinfo {author} {\bibfnamefont {P.~K.~S.}\
  \bibnamefont {Vaudrevange}},\ }\href {\doibase 10.1007/JHEP02(2020)045}
  {\bibfield  {journal} {\bibinfo  {journal} {JHEP}\ }\textbf {\bibinfo
  {volume} {02}},\ \bibinfo {pages} {045} (\bibinfo {year} {2020})},\ \Eprint
  {http://arxiv.org/abs/2001.01736} {arXiv:2001.01736 [hep-ph]} \BibitemShut
  {NoStop}%
\bibitem [{\citenamefont {Ishimori}\ \emph {et~al.}(2010)\citenamefont
  {Ishimori}, \citenamefont {Kobayashi}, \citenamefont {Ohki}, \citenamefont
  {Shimizu}, \citenamefont {Okada},\ and\ \citenamefont
  {Tanimoto}}]{Ishimori:2010au}%
  \BibitemOpen
  \bibfield  {author} {\bibinfo {author} {\bibfnamefont {H.}~\bibnamefont
  {Ishimori}}, \bibinfo {author} {\bibfnamefont {T.}~\bibnamefont {Kobayashi}},
  \bibinfo {author} {\bibfnamefont {H.}~\bibnamefont {Ohki}}, \bibinfo {author}
  {\bibfnamefont {Y.}~\bibnamefont {Shimizu}}, \bibinfo {author} {\bibfnamefont
  {H.}~\bibnamefont {Okada}}, \ and\ \bibinfo {author} {\bibfnamefont
  {M.}~\bibnamefont {Tanimoto}},\ }\href {\doibase 10.1143/PTPS.183.1}
  {\bibfield  {journal} {\bibinfo  {journal} {Prog. Theor. Phys. Suppl.}\
  }\textbf {\bibinfo {volume} {183}},\ \bibinfo {pages} {1} (\bibinfo {year}
  {2010})},\ \Eprint {http://arxiv.org/abs/1003.3552} {arXiv:1003.3552
  [hep-th]} \BibitemShut {NoStop}%
\bibitem [{\citenamefont {Holthausen}\ \emph
  {et~al.}(2013{\natexlab{b}})\citenamefont {Holthausen}, \citenamefont {Lim},\
  and\ \citenamefont {Lindner}}]{Holthausen:2012wt}%
  \BibitemOpen
  \bibfield  {author} {\bibinfo {author} {\bibfnamefont {M.}~\bibnamefont
  {Holthausen}}, \bibinfo {author} {\bibfnamefont {K.~S.}\ \bibnamefont {Lim}},
  \ and\ \bibinfo {author} {\bibfnamefont {M.}~\bibnamefont {Lindner}},\ }\href
  {\doibase 10.1016/j.physletb.2013.02.047} {\bibfield  {journal} {\bibinfo
  {journal} {Phys. Lett. B}\ }\textbf {\bibinfo {volume} {721}},\ \bibinfo
  {pages} {61} (\bibinfo {year} {2013}{\natexlab{b}})},\ \Eprint
  {http://arxiv.org/abs/1212.2411} {arXiv:1212.2411 [hep-ph]} \BibitemShut
  {NoStop}%
\bibitem [{\citenamefont {Ge}\ \emph {et~al.}(2012)\citenamefont {Ge},
  \citenamefont {Dicus},\ and\ \citenamefont {Repko}}]{Ge:2011qn}%
  \BibitemOpen
  \bibfield  {author} {\bibinfo {author} {\bibfnamefont {S.-F.}\ \bibnamefont
  {Ge}}, \bibinfo {author} {\bibfnamefont {D.~A.}\ \bibnamefont {Dicus}}, \
  and\ \bibinfo {author} {\bibfnamefont {W.~W.}\ \bibnamefont {Repko}},\ }\href
  {\doibase 10.1103/PhysRevLett.108.041801} {\bibfield  {journal} {\bibinfo
  {journal} {Phys. Rev. Lett.}\ }\textbf {\bibinfo {volume} {108}},\ \bibinfo
  {pages} {041801} (\bibinfo {year} {2012})},\ \Eprint
  {http://arxiv.org/abs/1108.0964} {arXiv:1108.0964 [hep-ph]} \BibitemShut
  {NoStop}%
\bibitem [{\citenamefont {Grimus}(2013)}]{Grimus:2013rw}%
  \BibitemOpen
  \bibfield  {author} {\bibinfo {author} {\bibfnamefont {W.}~\bibnamefont
  {Grimus}},\ }\href {\doibase 10.1088/0954-3899/40/7/075008} {\bibfield
  {journal} {\bibinfo  {journal} {J. Phys. G}\ }\textbf {\bibinfo {volume}
  {40}},\ \bibinfo {pages} {075008} (\bibinfo {year} {2013})},\ \Eprint
  {http://arxiv.org/abs/1301.0495} {arXiv:1301.0495 [hep-ph]} \BibitemShut
  {NoStop}%
\bibitem [{\citenamefont {Lavoura}\ and\ \citenamefont
  {Ludl}(2014)}]{Lavoura:2014kwa}%
  \BibitemOpen
  \bibfield  {author} {\bibinfo {author} {\bibfnamefont {L.}~\bibnamefont
  {Lavoura}}\ and\ \bibinfo {author} {\bibfnamefont {P.~O.}\ \bibnamefont
  {Ludl}},\ }\href {\doibase 10.1016/j.physletb.2014.03.001} {\bibfield
  {journal} {\bibinfo  {journal} {Phys. Lett. B}\ }\textbf {\bibinfo {volume}
  {731}},\ \bibinfo {pages} {331} (\bibinfo {year} {2014})},\ \Eprint
  {http://arxiv.org/abs/1401.5036} {arXiv:1401.5036 [hep-ph]} \BibitemShut
  {NoStop}%
\bibitem [{\citenamefont {Fonseca}\ and\ \citenamefont
  {Grimus}(2014)}]{Fonseca:2014koa}%
  \BibitemOpen
  \bibfield  {author} {\bibinfo {author} {\bibfnamefont {R.~M.}\ \bibnamefont
  {Fonseca}}\ and\ \bibinfo {author} {\bibfnamefont {W.}~\bibnamefont
  {Grimus}},\ }\href {\doibase 10.1007/JHEP09(2014)033} {\bibfield  {journal}
  {\bibinfo  {journal} {JHEP}\ }\textbf {\bibinfo {volume} {09}},\ \bibinfo
  {pages} {033} (\bibinfo {year} {2014})},\ \Eprint
  {http://arxiv.org/abs/1405.3678} {arXiv:1405.3678 [hep-ph]} \BibitemShut
  {NoStop}%
\bibitem [{\citenamefont {Esteban}\ \emph {et~al.}(2020)\citenamefont
  {Esteban}, \citenamefont {Gonzalez-Garcia}, \citenamefont {Maltoni},
  \citenamefont {Schwetz},\ and\ \citenamefont {Zhou}}]{Esteban:2020cvm}%
  \BibitemOpen
  \bibfield  {author} {\bibinfo {author} {\bibfnamefont {I.}~\bibnamefont
  {Esteban}}, \bibinfo {author} {\bibfnamefont {M.~C.}\ \bibnamefont
  {Gonzalez-Garcia}}, \bibinfo {author} {\bibfnamefont {M.}~\bibnamefont
  {Maltoni}}, \bibinfo {author} {\bibfnamefont {T.}~\bibnamefont {Schwetz}}, \
  and\ \bibinfo {author} {\bibfnamefont {A.}~\bibnamefont {Zhou}},\ }\href
  {\doibase 10.1007/JHEP09(2020)178} {\bibfield  {journal} {\bibinfo  {journal}
  {JHEP}\ }\textbf {\bibinfo {volume} {09}},\ \bibinfo {pages} {178} (\bibinfo
  {year} {2020})},\ \Eprint {http://arxiv.org/abs/2007.14792} {arXiv:2007.14792
  [hep-ph]} \BibitemShut {NoStop}%
\end{thebibliography}%

\end{document}